\documentclass[11pt,twoside]{article}
\usepackage{newpasp,epsf}
\def\edcomment#1{\iffalse\marginpar{\raggedright\sl#1\/}\else\relax\fi}
\reversemarginpar

\begin{document}
\title{Nonthermal Emission in Radio Galaxies from Simulated Relativistic 
Electron Transport in 3D MHD Flows}
\author{I. L. Tregillis, T. W. Jones}
\affil{School of Physics and Astronomy, University of Minnesota, 116 Church
St. S.E., Minneapolis, MN 55455, USA}
\author{Dongsu Ryu}
\affil{Department of Astronomy and Space Science, Chungnam National 
University, Daejon, 305-764, Korea}

\begin{abstract}
We perform a series of so-called ``synthetic observations'' on a set 
of 3D MHD jet simulations which explicitly include energy-dependent
transport of relativistic electrons, as described in the companion 
paper by Jones, Tregillis, \& Ryu.  Analyzing them in light of the 
complex source dynamics and energetic particle distributions described 
in that paper,  we find that the standard model for radiative aging in 
radio galaxies does not always adequately reflect the detailed source 
structure.
\end{abstract}

\section{Introduction}
	Jones, Ryu, \& Engel (1999) presented the first multidimensional 
MHD simulations to include explicitly time-dependent transport 
of the relativistic electrons responsible for the nonthermal emission 
observed from extragalactic radio
sources. The companion paper by Jones, Tregillis, \& Ryu (2000; paper I)
in this proceedings describes results from the extension of that work to 
three dimensions via exploratory simulations designed to probe the 
relationships between the source dynamics and the spatial and energy 
distributions of nonthermal electrons.  

	We compute self-consistent emission properties for the simulated
sources, thereby producing the first ``synthetic observations'' of their 
kind, including synchrotron surface brightness and spectral index maps.  
This paper briefly summarizes some of these synthetic observations.
A full report is in preparation.

\section{Synthetic Observation Methods}
	In every zone of the computational grid we compute an approximate,
self-consistent synchrotron emissivity as described in Jones et al. (1999).
This calculation is performed in the rest frame of the simulated radio galaxy,
with the appropriate redshift correction made for the observation frame.  It 
also explicitly takes into account the three-dimensional magnetic field
geometry.  Note that because we
are explicitly calculating the momentum distribution of nonthermal electrons,
we obtain the local synchrotron 
spectral index $\alpha$ directly.  Previously, spectral indices in synthetic 
observations from purely MHD simulations had to be included in an ad hoc 
fashion (Matthews \& Scheuer 1990; Clarke 1993). 

	Surface brightness maps for the 
optically thin emission are then produced via raytracing through the 
computational grid to perform line-of-sight integrations, thereby projecting 
the source on the plane of the sky at any arbitrary orientation.  
The synthetic observations can be imported into any standard image 
analysis package and subsequently analyzed like real observations; the 
analysis here was performed using the \textsc{MIRIAD} and \textsc{KARMA}
(Gooch 1995) packages.  
For example, it is a straightforward matter to construct spectral-index 
maps from a set of observations over a range of frequencies.  
To make this exercise as realistic as possible we place the 
simulated object at an appropriate luminosity distance, set to $100$ Mpc 
for the observations included here. Because our primary interest here is 
in identifying general trends, the observations are presented at their 
full resolution with very high dynamic range, although it is 
straightforward to convolve the images down to lower resolution before 
making comparisons to true observations.  

\section{Discussion}
	The dynamical effects outlined in the paper I have a profound 
impact on the nonthermal electron populations in these simulated sources.  
For the purpose of contrasting two extreme cases, here we will consider only
the ``injection'' and ``strong-cooling'' electron-transport models (models 
2 and 3 in paper I).  

\begin{figure}
\plotone{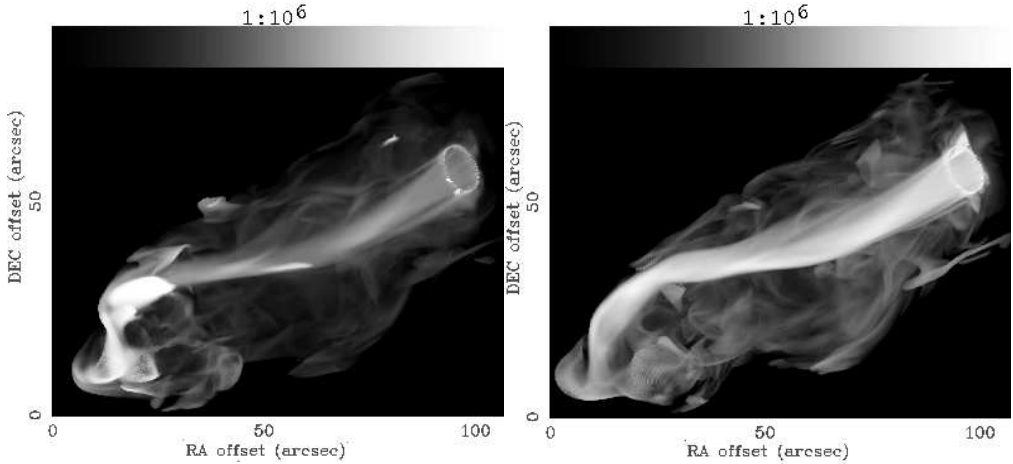}
\caption{1.4 GHz synchrotron surface brightness maps. The ``injection'' model
is on the left and ``strong-cooling'' on the right.}
\end{figure}

	Figure 1 shows synthetic synchrotron surface brightness
maps computed at 1.4 GHz, corresponding to a time when the jet has propagated
about 30 jet radii.  Consider first the ``injection'' transport model.  
A jet, hotspot complex, and lobe are all readily visible.  The bright ring 
in the upper right is the orifice where the jet enters the computational 
grid.  

	The apparent hotspot complex consists of a bright, compact
primary hotspot and a weaker, more diffuse secondary hotspot slightly below
it.  This secondary hotspot is a ``splatter spot'' 
(Cox, Gull, \& Scheuer 1991; Lonsdale \& Barthel 1986); we also see 
``dentist's drill'' type hotspots (Scheuer 1982) at other times.

In the case shown here, the bright primary hotspot does  
correspond to a strong shock at the jet terminus, although
this is often not the case.  
The synchrotron emissivity jumps by a factor of nearly $10^{4}$ across this
shock in the injection model because the model assumes almost no 
relativistic electrons enter with the jet flow. Rather, they are supplied
through shock injection. The magnetic pressure increases by merely a 
factor of $2.5$.  Thus, we have an example of a prominent
hotspot where the brightness is primarily the result of enhanced particle
populations rather than dramatic magnetic field growth.

	Contrast this with the situation in the ``strong cooling'' transport
model.  There, the
jet core dominates the emission, because  the
entire nonthermal particle population enters the grid with the jet.  Although
there is a minor brightness enhancement, we no longer see such a dramatic 
hotspot at the location of the terminal shock, despite the fact that these 
two models are dynamically identical.	Virtually all of the brightness
variations in this model are signaling variations in the underlying magnetic
field structure.

\begin{figure}
\plotone{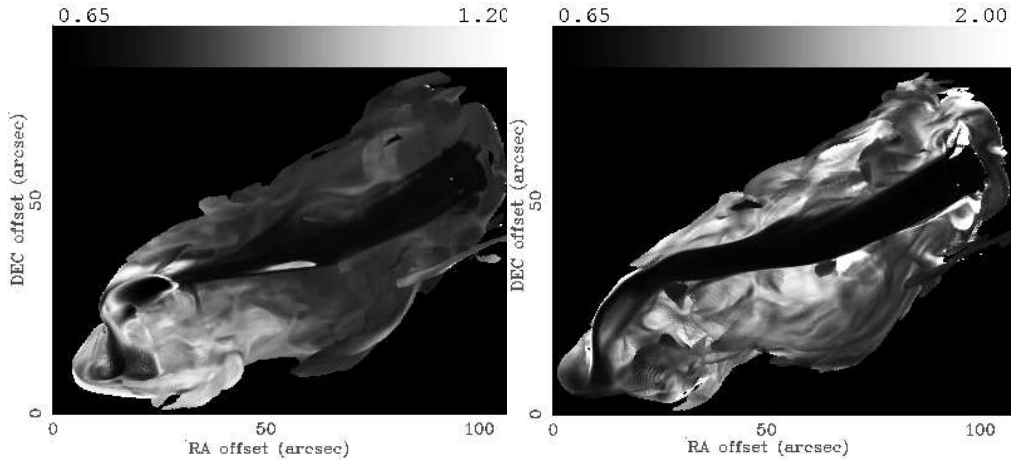}
\caption{Two-frequency spectral index maps computed from surface 
brightness at 1.4 and 5.2 GHz.  The ``injection'' model is on the left
and ``strong-cooling'' on the right.}
\end{figure}

	These models also give rise to fascinating spectral index maps.
Figure 2 shows two-frequency spectral index maps computed from
surface brightness maps at 1.4 and 5.2 GHz.  Once again considering the 
injection model first, we see that
the jet, hotspot, and lobe are easily identifiable.  The jet has a synchrotron
spectral index of 0.7, consistent with the momentum index of 4.4
for the particles sent down the jet.  The primary hotspot is slightly
flatter than this, in accordance with the presence of a strong shock.  
The lobe spectrum is somewhat peculiar, however, with material in the head
region steeper than that towards the core; this is quite the opposite from
what would be expected based on the standard paradigm for radio galaxy 
aging.  This surprising result is an excellent example of the impact of
the shock-web complex described in paper I.  Injection at the numerous 
weak shocks spread 
throughout the head region increases the steep-spectrum particle populations
enough that they can temporarily dominate the emission, a task made easier
by the relatively small flatter population sent down the jet in this model.  
The lobe emission becomes less steep as the two populations begin to mix
in the backflow, since radiative losses are minimal here.  Since injection
acts to create convex spectra, the emission is flattened even further as
the emitting particles diffuse into the large low-field volumes in the
cocoon.

	The complicated spectral index map from the strong cooling model 
vividly
demonstrates the extreme variability of the magnetic field in these sources.
Strong fluctuations in the field make it possible to create distinct
regions within the same source between which the effective cooling rate
differs by orders of magnitude.  Also, these sources are rather young in terms
of their histories, so individual events in the source evolution can still have
a discernible effect upon the overall appearance at these times.  For
instance, the region of moderately-steep material located below the jet
midway along the cocoon can be traced to a particular instance where 
shearing flows generated at the jet tip lead to a dramatic but transient
enhancement of the local magnetic field.  

	This work is supported by the U. S. National Science 
Foundation and the University of Minnesota Supercomputing Institute, 
and in Korea by KOSEF.


\begin{references}
\reference
Clarke, D. A. 1993, `3-D MHD simulations of extragalactic jets' in Lecture Notes in Physics Vol. 421, Jets in Extragalactic Radio Sources, ed. H.-J. R\"oser \& K. Meisenheimer (Berlin: Springer), 243--252

\reference
Cox, C. I., Gull, S. F., \& Scheuer, P. A. G. 1991, `Three-dimensional simulations of the jets of extragalactic radio sources', \mnras, 252, 558--585

\reference
Gooch, R. E. 1995, `Karma: a visualisation test-bed' in ASP Conf. Ser. Vol. 152, Astronomical Data Analysis Software and Systems V, ed. G. H. Jacoby \& J. Barnes, (San Francisco: ASP), 80--83

\reference
Jones, T. W., Tregillis, I. L., Ryu, Dongsu 2001, `3D MHD simulations of relativistic electron acceleration and transport in radio galaxies', these proceedings (paper I)

\reference
Jones, T. W., Ryu, Dongsu, \& Engel, Andrew 1999, `Simulating electron transport and synchrotron emission in radio galaxies: shock accleration and synchrotron aging in axisymmetric flows', \apj, 512, 105--124

\reference
Lonsdale, C. J., \& Barthel, P. D. 1986, `Double hotspots and flow redirection in the lobes of powerful extragalactic radio sources', \aj, 92, 12--22

\reference
Matthews, A. P., \& Scheuer, P. A. G. 1990, `Models of radio galaxies with tangled magnetic fields - I. calculation of magnetic field transport, Stokes parameters and synchrotron losses', \mnras, 242, 616--622

\reference 
Scheuer, P. A. G. 1982, `Morphology and power of radio sources', in IAU Symp. 97, Extragalactic Radio Sources, ed. D. S. Heeschen \& C. M. Wade (Dordrecht: Reidel), 163--165
\end{references}
\end{document}